\documentstyle[preprint,aps]{revtex}
\begin{document}
\draft
\title{Particle physics bounds from the Hulse-Taylor binary}
\author{Subhendra Mohanty\footnote[1]{mohanty@iopb.ernet.in}
and Prafulla Kumar Panda\footnote[2]{prafulla@iopb.ernet.in}}
\address{Institute of Physics, Bhubaneswar-751005, India.}
\maketitle
\begin{abstract}
The orbital period of the binary pulsar PSR 1913+16 has been observed
to decrease at the rate of $2.40\times 10^{-12}$ s/s which
agrees with the prediction of the quadropole formula for gravitational
radiation to within one percent. The decrease in orbital period may also
occur by radiation of other massless particles like scalars and
pseudoscalar Nambu-Goldstone bosons. Assuming that this energy loss is less
than one percent of the gravitational radiation, we can establish bounds
on couplings of these particles to nucleons. For a scalar
nucleon coupling of the form $g_s~\phi~\bar\psi~\psi$ we find that $g_s<
3\times 10^{-19}$. From the radiation loss of massless Goldstone
bosons we establish the upper bound $\theta /f< 7.5\times 10^{-16}$ GeV$^{-1}$
on the QCD vacuum angle $\theta$ and the scale $f$ at which the baryon
number symmetry can be broken spontaneously.
\end{abstract}

The Hulse-Taylor (H-T) binary consisting of the pulsar PSR 1913+16
orbiting around an unseen companion star provides firm evidence for
the existence of gravitational waves \cite{hul1}.
The observed loss of orbital period
agrees with the prediction from the quadropole formula of gravitation
radiation \cite{peter1} to within one percent. In this letter we compute the
orbital energy loss due to radiation of other massless particles
like scalars and pseudoscalar Nambu-Goldstone bosons. Massless
scalars which couple to nucleons arise in scalar-tensor theories
of gravity \cite{brans}, as dilatons in theories with spontaneously broken
conformal symmetry \cite{fujii} and in string theories \cite{green}.
Assuming a generic scalar-nucleon
coupling ${\cal L}_s=g_s\phi_s~\bar\psi~\psi$ we find that the radiation
of $\phi_s$ particles from the H-T binary is less than 1\% of the
gravitational radiation loss if $g_s<3\times 10^{-19}$.
This gives an upper bound
$\alpha_B~=~g_s^2/4\pi Gm_n^2 \leq 1$ on the ratio of a long range
scalar mediated fifth force to the gravitational force between two
nucleons. This bound is not as
stringent as the bounds obtained from terrestrial fifth force
search experiments \cite{adel,stubbs} which give $\alpha_B$ in the
range of $10^{-3}~-~ 10^{-6}$.

In theories with a spontaneously broken global symmetry like the baryon
number or the lepton number we have massless Nambu-Goldstone bosons (NGB)
which have a generic coupling ${\cal L}_p=(m/f)
\phi_p~\bar\psi i \gamma_5 \psi$ where $m$ is the fermion mass and $f$
is the scale of the global symmetry breaking \cite{moha1}.
The pseudoscalar field
of a macroscopic source adds up coherently only if the spins of the
constituents are polarised. It was observed by Chang, Mohapatra and
Nussinov \cite{chang} and Barbieri et al \cite{barb}
that the CP violating operator $\theta G\tilde G$ in the
QCD sector induces a coupling ${\cal L}=(\theta /f)
(m_um_d/(m_u+m_d)) \phi_p\bar\psi \psi$ between the
NGB $\phi_p$ and nucleons. This coupling give rise to a $1/r$
type long range force and the NGB field of the constituent nucleons
a macroscopic test body add up coherently even when their spins are randomly
aligned. From the constraints
on the energy carried away by the radiation of NGBs from the H-T
binary we obtain the upper bound $(\theta /f) < 7.5\times~10^{-16}$ GeV$^{-1}$.
This can be compared with the
separately established bounds $\theta < 10^{-9}$ (from the measurement of
the neutron electric dipole moment \cite{nedme,nedmt}) and $f> 10^8$ GeV
(from the cooling rate of helium stars \cite{dicus}).
The rate of energy loss by scalar particle emission is $\propto ~\Omega^4$
(where $\Omega$ is the orbital frequency). Observations of binary systems
with faster orbital frequencies can be used to put more stringent bounds
on couplings considered in this paper.

Finally we compute the energy loss by the radiation of neutrino pairs
from the constituent neutrons of the H-T binary. We find that for the
neutral current coupling ${\cal L}_\nu=(1/{\sqrt 2})G_F n(x)\bar\nu \nu$, the
energy radiated by neutrino pair emission is suppressed by the phase factor
and is negligibly smaller than the gravitational radiation.
Therefore if experimentally one observes a discrepancy
between the observed period loss of the
binary orbit and the prediction from gravitational radiation formula, it
would be a signal of new kind of massless particle radiation and
signal of new physics beyond the standard model.

\noindent {\it Gravitational radiation :}
We use the Feynman rules of linearised quantum gravity
\cite{feyn,kibble1,isham1,boul1} to compute
the gravitational radiation from the Hulse-Taylor
binary system. Assuming a universal graviton-matter coupling
${\cal L}_I= \sqrt{8\pi G} ~ h^{\mu\nu}~T_{\mu\nu}$ we find
that the energy loss by gravitational brehmstrahlung at tree level
agrees with the Peter-Mathew expression \cite{peter1}
for classical gravitational wave radiation from binary system.

The effective Lagrangian for the graviton matter interaction is
\begin{equation}
{\cal L}=-\frac{1}{4}{\tilde h}^{\mu\nu}\Box {\tilde h}_{\mu\nu}
+ \kappa ~{\tilde h}^{\mu\nu}~ T_{\mu\nu}
\end{equation}
where the graviton field $ h_{\mu\nu} $ is a perturbation
of the metric $g_{\mu\nu}= \eta_{\mu\nu}
+ \kappa h_{\mu\nu}$ to the first order in $\kappa=\sqrt{8\pi G}$. We have
made the harmonic gauge choice,
 $\partial_\mu h^\mu~ _\lambda =(1/2)\partial_\nu h^\mu_\mu$ and defined
${\tilde h}_{\mu\nu}=h_{\mu\nu}-\frac{1}{2}\eta_{\mu\nu}h^\lambda~_\lambda$
The universal coupling of ${\tilde h}^{\mu\nu}$ with matter at the tree
level is a reflection of the equivalence principle of classical gravity.
The rate of graviton emission is given by
\begin{equation}
d\Gamma=\kappa^2\sum_{\lambda=1}^2 | T_{\mu\nu}(k')~
{\tilde \epsilon}_{(\lambda)}^{\mu\nu}~(k)|^2 ~2\pi~
\delta(\omega-\omega')~
\frac{d^3k}{(2\pi)^3}~\frac{1}{2\omega}
\label{dgama1}
\end{equation}
where $T_{\mu\nu}(k')$ and $\tilde\epsilon^{\mu\nu}_{\lambda}(k)$ are the
Fourier transforms of $T_{\mu\nu}(x)$ and $h^{\mu\nu}(x)$ and $\lambda$
denotes the polarisation of the emitted gravitons which is summed over.
Using the harmonic gauge condition
$k^\mu ~(\epsilon_{\mu\nu}-\frac{1}{2}~
\eta_{\mu\nu}\epsilon^\lambda~_\lambda)=0,$
we can set $\epsilon_{\mu 0}=0$, $\epsilon^i~_i=0$, $k^i\epsilon_{ij}=0$
we have ${\tilde \epsilon}_{\mu\nu}=\epsilon_{\mu\nu}$ equation
(\ref{dgama1}) can be written as
\begin{equation}
d\Gamma=\frac{1}{2}~\frac{\kappa^2}{(2\pi)^2}\sum_{\lambda=1}^2
{\epsilon_{(\lambda)}^{ij}}^* (k)~
{\epsilon_{(\lambda)}^{lm}}(k)~ T_{ij}^*(\omega')~ T_{lm}(\omega')~ \omega
{}~\delta (\omega -\omega')~ d\omega ~d\Omega_k
\end{equation}
Using the relation
\begin{equation}
\int d\Omega_k\sum_{\lambda=1}^2
{\epsilon_{(\lambda)}^{ij}}^* (k)~{\epsilon_{(\lambda)}^{lm}}(k)
=\frac{8\pi}{5}\left[\frac{1}{2}\Big(\delta_{il}\delta_{jm}
+\delta_{im}~\delta_{jl}\Big)
-\frac{1}{3}\delta_{ij}\delta_{lm}\right]
\end{equation}
We have the graviton emission rate
\begin{equation}
d\Gamma=\frac{\kappa^2}{5\pi}\Big[ T_{ij}(\omega') T_{ji}^* (\omega')
-\frac{1}{3} |T~^i ~_i(\omega')|^2\Big]\omega~\delta(\omega-\omega') ~d\omega
\end{equation}
and the rate of energy loss by the graviton radiation is
\begin{equation}
\frac{dE}{dt}=\int\frac{\kappa^2}{5\pi}~\omega^2
\Big[T_{ij}(\omega') T_{ji}^*(\omega') -\frac{1}{3} |T~^i ~_i(\omega')
|^2\Big]~\delta(\omega-\omega')~d\omega
\label{dedt}
\end{equation}
For a binary system of stars with masses $m_1$, $m_2$ in an eliptical orbit
around the center of mass, the stress tensor is given by
\begin{equation}
T_{\mu\nu}(x)=M \delta^3 (\vec x-\vec y(t)) U_\mu U_\nu
\label{eq:tmn1}
\end{equation}
where $M~=~m_1 m_2/(m_1+m_2)$ is the reduced mass and $U_\mu=~\gamma
(1,\dot{x},
\dot{y}, \dot{z})$ is the four velocity of the
reduced mass in the elliptical orbit.
Assuming an orbit in the $x-y$ plane with  coordinates
$\vec y(t)=~(d~ cos\theta ,~d~ sin \theta)$,
the relative distance $d$ and the angular velocity $\dot{\theta}$ are
given by
\begin{equation}
d=\frac{a(1-e^2)}{1+e~cos\theta}
\label{d}
\end{equation}
\begin{equation}
\dot{\theta}=\left[\frac{G(m_1+m_2) ~a~(1-e^2)}{d^2}\right]^{1/2}
\label{thd}
\end{equation}
where $a$ is the semimajor axis and $e$ the eccentricity of the
eliptical orbit.
Substitution (\ref{d}) and (\ref{thd}) in (\ref{eq:tmn1})
and taking the fourier transform
we obtain the stress tensor $T_{ij} (\omega'=n\Omega)$ for the Kepler
orbit [2] in terms of the $n$ harmonics of the fundamental frequency
$\Omega=(G(m_1+m_2)~a^{-3})^{1/2}$,
\begin{mathletters}
\begin{equation}
T_{xx}(\omega')=-\frac{{\omega'}^2}{4}~M ~ a^2 \frac{2}{n}[ J_{n-2}(ne)
-2e~J_{n-1}(ne)+2e~ J_{n+1}(ne)-J_{n+2}(ne)]
\label{txx}
\end{equation}
\begin{equation}
T_{yy}(\omega')=\frac{{\omega'}^2}{4}~M ~ a^2 \frac{2}{n}[ J_{n-2}(ne)
-2e~J_{n-1}(ne)+\frac{4}{n}~J_n(ne)+2e~ J_{n+1}(ne)-J_{n+2}(ne)]
\label{tyy}
\end{equation}
\begin{equation}
T_{xy}(\omega')=\frac{{\omega'}^2}{4i}~M ~ a^2
\frac{2}{n}~(1-e^2)^{1/2}[ J_{n-2}(ne) -2~J_{n}(ne)+J_{n+2}(ne)]
\label{txy}
\end{equation}
\end{mathletters}
 From the above equation we obtain
\begin{equation}
\Big[T_{ij}(\omega') T_{ji}^*(\omega')-\frac{1}{3}
|T~^i ~_i(\omega')|^2\Big]= ~ M^2 ~a^4 ~{\omega'}^4 ~g(n,e)
\label{tt}
\end{equation}
where
\begin{eqnarray}
g(n,e)&=&\frac{1}{32~n^2}\Big\{ [ J_{n-2}(ne)-2e~J_{n-1}(ne)
+2e~J_{n+1}(ne)-J_{n+2}(ne)]^2 \nonumber\\ &+&(1-e^2)[ J_{n-2}(ne)-2~J_{n}(ne)
+J_{n+2}(ne)]^2+\frac{4}{3n^2}[J_n(ne)]^2\Big\}
\end{eqnarray}
Substituting (\ref{tt}) into (\ref{dedt}) we have
the rate of energy loss of a binary
system in an elliptical orbit as sum of radiation in the $n$ harmonics
of the fundamental frequency $\Omega$,
\begin{equation}
\frac{dE}{dt}=\frac{32 G}{5}~\sum_n(n~\Omega)^2
\cdot M^2~a^4~ (n~\Omega)^4~g(n,e)
\label{dedt1}
\end{equation}
Using the relation
$\sum_n^\infty n^6~g(n,e)=(1-e^2)^{-7/2}\left( 1+ \frac{73}{24} e^2
+\frac{37}{96}e^4\right)$ [from \cite{peter1}]
and substituting the values of $\Omega=0.2251\times 10^{-3}$
sec$^{-1}$, $m_1=1.42M_\odot$,
$m_2=1.4M_\odot$, $a=3.0813815$ lsec, $e=0.617127$ for the parameters of
the H-T binary, we find that the energy loss is
\begin{eqnarray}
\frac{dE}{dt}&=&\frac{32}{5}\cdot G\cdot \Omega^6\cdot
\left(\frac{m_1m_2}{m_1+m_2}\right)^2 ~ a^4~
(1-e^2)^{-7/2}\left( 1+ \frac{73}{24} e^2 +\frac{37}{96}e^4\right)
\nonumber\\ &=&3.2\times 10^{33}~ \mbox{erg/sec}
\end{eqnarray}
The time period of the elliptical orbit depends upon the energy $E$,
so energy loss leads to a change of the time period of the orbit at the rate
\begin{eqnarray}
\frac{dP_b}{dt}&=&-6\pi G^{-3/2}(m_1 m_2)^{-1}
(m_1+m_2)^{-1/2} a^{5/2} \left(\frac{dE}{dt}\right)\nonumber\\
&=&\frac{192\pi}{5}G^{5/3} \Omega^{5/3}\frac{m_1 m_2}
{(m_1+m_2)^{1/3}}(1-e^2)^{-7/2}\left( 1+ \frac{73}{24} e^2+\frac{37}{96}e^4
\right)
\label{dpbdt1}
\end{eqnarray}
which is the same expression as obtained by Peter and Mathews \cite{peter1}
for the classical
gravitational radiation from binary systems. We find therefore that the
tree level quantum gravity calculation agrees with the clasical results.
For the Hulse - Taylor binary, the expression (\ref{dpbdt1})
yields the orbital period decceleration due to the gravitational radiation
$ \dot{P}_b=-2.403\pm 0.002\times 10^{-12}$
which agrees with the observed value from the Hulse - Taylor binary
\cite{hul1} $ \dot{P}_b (observed)=-2.40\pm 0.09\times 10^{-12}$
to with in 1\%. Energy loss by emission of other massless
particles should be within 1\% of energy loss by gravitational
radiation and that can be used to put bounds on the couplings
of the various massless scalar and pseudoscalar particles which
arise in particle physics models.

\noindent{\it Massless scalar radiation:}
We assume a coupling between massless scalar fields $\phi_s$
and the baryons of the form
\begin{equation}
{\cal L}_s=g_s~ \phi_s~ \bar\psi~\psi
\label{eq:ls1}
\end{equation}
which for a macroscopic baryon source can be written as
\begin{equation}
{\cal L}_s=g_s~\phi_s~n(x)
\label{eq:ls2}
\end{equation}
where $n(x)$ is the baryon number density. A neutron star with
radius $\sim 10$ km can be regarded as a point source since the
Compton wavelength of the radiation $\sim \Omega^{-1}=10^9$ km is much
larger than the dimension of the source. The baryon number density $n(x)$
for the binary stars (denoted by $a=1,2$) may be written as
\begin{equation}
n(x)=\sum_{a=1,2}~N_a~\delta^3(\vec x-\vec x_a(t))
\end{equation}
where $N_a\sim 10^{57}$ is the total number of baryons in the neutron star
and $\vec x_a(t)$ represents the Keplerian orbit of the binary stars.
For the coupling (\ref{eq:ls2}) and the source (18) the
rate of scalar particles emitted from the neutron star in orbit with
frequency $\Omega$ is
\begin{equation}
d\Gamma=|n(\omega)|^2 (2\pi)~\delta(\omega-\omega')\frac{d^3\omega'}
{(2\pi)^3~2\omega'}
\end{equation}
the rate of energy loss by massless scalar radiation is
\begin{equation}
\frac{dE}{dt}=\int |n(\omega)|^2~\omega'~(2\pi)~\delta(\omega-\omega')
\frac{d^3\omega'}{(2\pi)^3~2\omega'}
\end{equation}
where $n(\omega)$ is the fourier expansion of the source density (18)
\begin{equation}
n(\omega)=~\frac{1}{2\pi}\int~e^{i\vec k\cdot \vec x}~e^{-i\omega t}
\sum_{a=1,2}~N_a \delta^3(\vec x-\vec x_a(t))d^3x~ dt
\end{equation}
with $\omega=n\Omega$. Going over to the c.m. coordinates $\vec r~=~(x,y)$
by substituting $\vec x_1~= ~m_2\vec r /(m_1+m_2)$,
$\vec x_2~= ~-m_1\vec r /(m_1+m_2)$ we have
\begin{equation}
n(\omega)=~(N_1+N_2)~\delta(\omega)~+\left(\frac{N_1}{m_1}-\frac{N_2}{m_2}
\right)~M~(~ik_x~x(\omega)~+~ik_y~y(\omega))+ O(\vec k,\vec r)^2
\end{equation}
where $(x(\omega),y(\omega))$ are the fourier components of the Kepler orbit
of the reduced mass in the c.m. frame given by (8)-(9).
The first term in $n(\omega)$ is a delta function which has vanishing
contribution to (20). The leading non-zero contribution
comes from the second term in (22). Substituting \cite{landau}
\begin{equation}
x(\omega)~=~\frac{2~a}{n}~J'_n(ne),\quad y(\omega)~=~\frac{2i~a~\sqrt{1-e}}{ne}
J_n(ne)
\end{equation}
in (22) we obtain the expression for $|n(\omega)|^2$ given by
\begin{equation}
|n(\omega)|^2~=~\frac{4}{3}\left[\left(\frac{N_1}{m_1}-
\frac{N_2}{m_2}\right)~M\right]^2a^2
\Omega^2~[{J'}^2_n(ne)+\frac{(1-e^2)}{e^2}
{}~J^2_n(ne)]
\end{equation}
where we have used the dispersion relation $k_x^2~=~k_y^2~=
{}~\frac{1}{3}(n\Omega)^2$. Substituting (24) in (20)
we have the rate of energy loss by massless scalars
\begin{equation}
\frac{dE}{dt}~=~\frac{2}{3\pi}\left[\left(\frac{N_1}{m_1}
-\frac{N_2}{m_2}\right)~M~g_s\right]^2
a^2\Omega^4~\sum_n~ n^2~[{J'}^2_n(ne)+\frac{(1-e^2)}{e^2}~J^2_n(ne)]
\end{equation}
The mode sum can be carried out using the Bessels function series
formulas given in reference \cite{peter1} to give
$\sum_n~ n^2~[{J'}^2_n(ne)+(1-e^2)e^{-2}~J^2_n(ne)]~=~(1/4)~
(2+e^2)(1-e^2)^{-5/2}$.
The energy loss (25) in terms of the orbital parameters $\Omega$, $a$ and
$e$ is given by
\begin{equation}
\frac{dE}{dt}~=~ \frac{1}{3\pi}~\left[\left(\frac{N_1}{m_1}-
\frac{N_2}{m_2}\right)~M~g_s\right]^2~\Omega^4 ~ a^2~
\frac{(1+e^2/2)}{(1-e^2)^{5/2}}
\end{equation}
Since $N_a~m_n=~m_a-\epsilon_a$ where $\epsilon_a=\frac{Gm_a^2}{R_a}$
is the gravitational binding energy and $m_n$ the neutron mass, the factor
$\left(\frac{N_1}{m_1}-\frac{N_2}{m_2}\right)=G\left(\frac{m_1}{R_1}-
\frac{m_2}{R_2}\right)$. For the H-T binary $m_1-m_2\simeq 0.02 $$M_\odot$
and $R_a\sim 10$ km, therefore
$\left(\frac{N_1}{m_1}-\frac{N_2}{m_2}\right)\simeq 3\times 10^{-3}$
GeV$^{-1}$.

For the H-T binary the rate of energy loss terms out to be
\begin{equation}
\frac{dE}{dt}~=~ g_s^2~ \times~9.62~\times ~10^{67}~\mbox{ergs/sec}
\end{equation}

Assuming that this is less than 1\% of the gravitational energy loss
i.e. $dE/dt\leq 10^{31}$ ergs/sec, we obtain
gives an upper bound on scalar nucleon coupling
\begin{equation}
g_s <~3~\times~ 10^{-19}.
\end{equation}
Exchange of massless scalar between two nucleons
gives rise to spin independent
fifth force with the static potential $V_{ss}(\vec r)=-~g_s^2/4\pi r$,
which shows that ratio of the fifth force to the gravitational force
between two nucleons is
$\alpha_B~=~\frac{g_s^2}{4\pi Gm_n^2} \leq 1.$
This is less stringent than the bound
$\alpha_B ~\sim ~10^{-3}-10^{-6}$
obtained from terrestrial fifth force search experiments
\cite{adel,stubbs,moha1}.

\noindent{\it Nambu-Goldstone boson radiation:}
Massless pseudoscalar particles arise
as Nambu-Goldstone bosons (NGB) when
some global symmetry is broken spontaneously \cite{nambu1,nambu2,nambu3}.
We  consider the coupling of massless NGBs $\phi_p$ to baryons,
\begin{equation}
{\cal L}_p~=~\frac{1}{f}~(\partial_\mu\phi_p)\bar\psi \gamma^\mu\gamma^5
\psi~=~\frac{m}{f}~\phi_p (\bar\psi i \gamma_5 \psi )
\label{eq:lp1}
\end{equation}
(where the last equality holds for fermions on mass shell)
which arise in many particle physics models
where a global symmetry is broken spontaneously at some scale
$f$ giving rise to Nambu-Goldstone bosons. This coupling gives rise
to a spin dependent long-range force
\cite{moody} $V(r) \sim (1/f)^2 (1/r^3)(\sigma_1\cdot
\sigma_2 - 3(\sigma_1\cdot \hat r)(\sigma_2\cdot \hat r))$.
The pseudoscalar coupling in the first order in $g_p$ is spin dependent
and the field of a macroscopic body with randomly oriented constituents
averages to zero.
The radiation of NGB by the $N$ constituent particles of the macroscopic
system will be in coherent which means that the energy radiated by $N$
particles will be $N$ times the single particle energy loss. This is
different from scalar radiation which being coherent scales as $N^2$ times
the single particle radiation.

It was observed by Chang, Mohapatra and Nussinov
\cite{chang} and Barbieri et al \cite{barb}
that if there is CP violation in the theory then $\phi_p$ can have both
pseudoscalar coupling as in (\ref{eq:lp1}) as well as scalar coupling
of the form (\ref{eq:ls1}). For example the CP violating
QCD term $\theta G \tilde G$ will induce an interaction between
$\phi_p$ and nucleons of the form
\begin{equation}
{\cal L}_p~=~\frac{\theta}{f}~\left(\frac{m_um_d}{m_u+m_d}\right)
\phi_p\bar \psi \psi
\label{eq:lp2}
\end{equation}
This scalar coupling gives rise to a long range $(V(r)\sim 1/r)$
potential which is spin independent so that $\phi_p$ field outside a
macroscopic object adds up coherently. The form of the interaction term
(\ref{eq:lp2}) is the same as in equation (\ref{eq:ls1})
so the same bound (28) holds for the dimensionless coupling
\begin{equation}
\frac{\theta}{f}\left(\frac{m_um_d}{m_u+m_d}\right) < 3\times~10 ^{-19}
\label{eq:bound1}
\end{equation}
which means that for $m_u=5$ MeV, $m_d=9$ MeV we have
$\theta /f < 7.5\times 10^{-16}$ GeV$^{-1}$. This can be compared with the
separate bounds $\theta < 10^{-9}$
(obtained from neutron edm \cite{nedme,nedmt})
and $f> 10^{8}$ GeV (obtained from the
cooling rate of helium stars \cite{dicus}). The bound (\ref{eq:bound1})
holds for models where the mass of the Nambu-Goldstone boson is smaller than
the
frequency $\Omega\simeq 10^{-19}$ eV of the binary orbit. In axion models
\cite{axion} the axion acquires a mass due to QCD instantons which
ranges between $10^{-3}$ eV $-$ $10^{-6}$ eV,
so our bound does not hold for these model. In majorons
models \cite{nambu1,nambu2}
where the lepton number is broken spontaneously, the majorons which
arise as NGBs can remain massless and our bound would then hold.

\noindent {\it Neutrino radiation :}
The coupling of neutrinos to neutrons in the standard
model is via the weak neutral current and is given by
\begin{equation}
{\cal L}_\nu={\sqrt 2} G_F \frac{n(x)}{2}\bar\nu_L\gamma^0\nu_L
\label{eq:lnu1}
\end{equation}
where $n(x)$ is the number density of the neutrons which are the
source of the neutrino field. The radiation of neutrinos from the H-T
binary with $n(x)=N~\delta^3(x-x(t))$ carries aways energy at the rate
\begin{equation}
\frac{dE}{dt}\simeq \left( \frac{G_F^2 N^2}{4}\right)\Omega^6=
10^{-43} ~~ \mbox{ergs/sec}.
\end{equation}
The weak coupling of neutrinos to neutrons is much larger
than the coupling of the gravitons to matter, however since the emission of
neutrinos occurs in pairs, the phase space suppression makes
the energy radiated by neutrino emission negligible compared to graviational
radiation.

The period loss in the H-T system has been determined by measuring the time of
periastron over a period of almost 19 years. The accuracy of the measured value
of period loss increases quadratically with time. If in the course
of observation one finds a significant discrepancy between the
observed value of period loss and the prediction of the gravitational
quadropole formula, it would be a compelling signal
of physics beyond standard model.

\noindent{\it Note added:}  Astrophysical bounds from the Hulse-Taylor
binary have been considered in references \cite{will} and \cite{cliff}.
In \cite{will} the energy loss in Brans-Dicke gravity is derived and
in \cite{cliff} effect of scalar couplings on orbital parameters of the
H-T binary is studied.

\end{document}